\documentclass[cernpreprint,texlive=2011,txfonts,UKenglish]{latex/atlasdoc}
\pdfoutput=1
\usepackage[compress,comma,numbers]{natbib}
\usepackage[biblatex=false]{latex/atlaspackage}
\usepackage{latex/atlasphysics}
\graphicspath{{logos/}{figures/}}
\nolinenumbers
\AtlasTitle{Measurement of transverse energy--energy correlations in multi-jet 
events in $pp$ collisions at $\sqrt{s} = 7 \TeV$ using the ATLAS detector and 
determination of the strong coupling constant $\alpha_{\mathrm{s}}(m_\Zboson)$}

\author{The ATLAS Collaboration}

\PreprintIdNumber{CERN-PH-EP-2015-177}
\AtlasJournal{Phys. Lett. B}

\AtlasAbstract{
High transverse momentum jets produced in pp collisions at a centre of mass energy of $7 \TeV$
are used to measure the transverse energy--energy correlation function and its associated azimuthal
asymmetry. The data were recorded with the ATLAS detector at the LHC in the year 2011 and correspond to an integrated luminosity of
158 $\ipb$. The selection criteria demand the average transverse momentum of the two leading jets in an event to be larger than $250 \GeV$. The data at detector level are well described by Monte Carlo event generators. They 
are unfolded to the particle level and compared with theoretical calculations at 
next-to-leading-order accuracy. The agreement between data and theory is good and 
provides a precision test of perturbative Quantum Chromodynamics at large momentum transfers. From
this comparison, the strong coupling constant given at the $\Zboson$ boson mass is determined to be
$\alpha_{\mathrm{s}}(m_\Zboson) = 0.1173 \pm 0.0010 \mbox{ (exp.) }^{+0.0065}_{-0.0026} \mbox{ (theo.)}$.
}

\newcommand\T{\rule{0pt}{2.ex}}
\newcommand\B{\rule[-1.2ex]{0pt}{0pt}}
\newcommand\TA{\rule{0pt}{2.2ex}}
\newcommand\TB{\rule{0pt}{2.3ex}}

\begin{document}
\maketitle
\tableofcontents
\clearpage

\section{Introduction}
\label{sec:intro}
The study of jet production at the LHC provides a quantitative test of  Quantum
Chromodynamics, QCD, at the highest momentum transfers. Theoretical calculations
for jet cross-sections in hadronic collisions have been carried out up to next-to-leading
order (NLO) accuracy in the strong coupling constant $\alpha_{\mathrm{s}}$ \cite{Catani,Nagy1,Nagy2}
and extensively compared with the data \cite{Aadxs1a,Aadxs1b,Aadxs2,atlas3jet,Chatrchyanxs1,Chatrchyanxs2,cms3jet}.
These calculations are valid for configurations with up to four jets in the final state.\\
Event shape variables have been measured in all major $e^+e^-$ experiments, as well as
in experiments at the electron--proton collider HERA. These studies were recently extended to hadron
colliders with measurements of the transverse thrust and the transverse minor
\cite{Banfi1,Banfi2} at the Tevatron \cite{Aaltonen} and the LHC \cite{Khachatryan,Aad}.\\
Energy--energy correlations (EEC), i.e.\ measurements of the energy-weighted angular
distributions of hadron pairs produced in $e^+e^-$ annihilation, were proposed in
Refs. \cite{Basham1,Basham2} as an alternative event shape variable not based on the
determination of the thrust principal axis \cite{brandt} or the sphericity tensor \cite{bjo}.
The EEC function and its asymmetry, AEEC, were subsequently 
calculated in $\mathcal{O}(\alpha_{\mathrm{s}}^2)$ \cite{Ali1982,Ali1984A,Richards}, and their measurements 
\cite{PLUTO,MARKII,MARKJ,CELLO,JADE,MAC,TASSO,TOPAZ,DELPHI,OPAL,ALEPH,L3,SLD}
have had significant impact on the precision 
tests of perturbative QCD and in the determination of the strong coupling constant 
in $e^+e^-$ annihilation experiments; a recent review is given in Ref.~\cite{Ali2010}. The EEC 
are by construction not affected by soft divergences, and as a consequence of this 
they are calculable at high orders.\\
The transverse energy--energy correlation function, TEEC, and its asymmetry, ATEEC, were proposed  
as the analogous variables at hadron collider experiments in Ref.~\cite{Ali1984B}, where 
predictions to leading order (LO) were also presented. The NLO 
corrections were calculated recently in Ref.~\cite{Ali2013} using \textsc{NLOJet++} \cite{Nagy1,Nagy2}. These calculations allow 
for a numerical determination of the NLO predictions for the TEEC and ATEEC, i.e.\ the coefficients of the second order polynomials in the strong
coupling constant. They are used in this paper for quantitative precision tests of QCD
including a determination of the strong coupling constant. The TEEC is defined as:
\begin{equation}
\hspace{-0.7cm}\frac{1}{\sigma}\frac{\mathrm{d}\Sigma}{\mathrm{d}(\cos\phi)} = \frac{1}{\sigma} \sum_{ij} \int \frac{\mathrm{d}\sigma}{\mathrm{d}\xti \mathrm{d}\xtj \mathrm{d}(\cos\phi)} \xti \xtj \mathrm{d}\xti \mathrm{d}\xtj,
\label{eq:defEEC}
\end{equation}
where the sum runs over all pairs of jets 
in the final state with azimuthal\footnote{ATLAS uses a right-handed
coordinate system with its origin at the nominal interaction point (IP) in the centre
of the detector and the $z$-axis along the beam pipe. The $x$-axis points from the IP
to the centre of the LHC ring, and the $y$-axis points upward. Cylindrical coordinates
$(r,\varphi)$ are used in the transverse plane, $\varphi$ being the azimuthal angle
around the beam pipe. The pseudorapidity is defined in terms of the polar angle
$\theta$ as $\eta=-\ln\tan(\theta/2)$.} angular difference $\phi = \Delta \varphi_{ij}$ and $\xti = \eti / \et$
is the transverse energy carried by jet $i$ in units of the sum of jet transverse energies $\et =\sum_i \eti$.
In order to cancel uncertainties that are constant over $\cos\phi \in [-1, 1]$, it 
is useful to define the azimuthal asymmetry of the TEEC (ATEEC) as
\begin{equation}
\frac{1}{\sigma}\frac{\mathrm{d}\Sigma^{{\rm asym}}}{\mathrm{d}(\cos\phi)}\equiv
\left.\frac{1}{\sigma}\frac{\mathrm{d}\Sigma}{\mathrm{d}(\cos\phi)}\right|_{\phi} -
\left.\frac{1}{\sigma}\frac{\mathrm{d}\Sigma}{\mathrm{d}(\cos\phi)}\right|_{\pi -\phi}.
\label{eq:defAEEC}
\end{equation}
This Letter presents a measurement of the TEEC and its associated asymmetry using high-energy jets.
\section{The ATLAS detector}
\label{sec:detector}
The ATLAS detector \cite{detector} is a multi-purpose particle physics detector with 
a forward-backward symmetric cylindrical geometry and a solid angle coverage of almost $4\pi$.\\
The inner tracking system covers the pseudorapidity range $|\eta|< 2.5$, and consists 
of a silicon pixel detector, a silicon microstrip detector, and, for $|\eta|<2.0$, a 
transition radiation tracker. It is surrounded by a thin 
superconducting solenoid providing a 2 $\mathrm{T}$ magnetic field along the beam 
direction. A high-granularity liquid-argon sampling electromagnetic calorimeter covers 
the region $|\eta|<3.2$. An iron/scintillator tile hadronic calorimeter provides 
coverage in the range $|\eta|<1.7$. The endcap and forward regions, spanning 
$1.5<|\eta|<4.9$, are instrumented with liquid-argon calorimeters for electromagnetic 
and hadronic measurements. The muon spectrometer surrounds the calorimeters. It consists 
of three large air-core superconducting toroid systems and separate trigger and 
high-precision tracking chambers providing accurate muon tracking for $|\eta|<2.7$.\\
The trigger system \cite{atlasTrigger} has three consecutive levels: level 1 (L1), 
level 2 (L2) and the event filter (EF). The L1 triggers are hardware-based and use 
coarse detector information to identify regions of interest, whereas the L2 triggers 
are software-based and perform a fast online data reconstruction. Finally, the EF 
uses reconstruction algorithms similar to the offline versions with the full 
detector granularity.
\section{Monte Carlo samples}
\label{sec:mc}
Multi-jet production in $pp$ collisions is represented by the convolution
of the production cross-sections for parton--parton scattering with the parton distribution functions. 
Monte Carlo (MC) generators differ in the approximations used to
calculate the underlying short-distance QCD process, in the way parton showers are built to
take into account higher-order effects and in the fragmentation scheme responsible for
long-distance effects. 
For this analysis, two different MC approaches are used, depending on 
whether the underlying hard process is considered to be $2\to2$ or multi-legged. The 
generated events are then processed with the ATLAS full detector simulation 
\cite{atlasim} based on \textsc{Geant4} \cite{geant}.\\
The baseline MC samples are generated using \textsc{Pythia 6.423} \cite{pythia}
with the matrix elements for the underlying $2 \to 2$ processes calculated
at LO using the \textsc{MRST2007LO*} parton distribution functions (PDF) \cite{pdf2}
and matched to transverse-momentum-ordered parton showers. The \textsc{AUET2B} tune
\cite{atlasTunes,atlasTunes2} is used to model the underlying event (UE) and the hadronisation
follows the Lund string model \cite{lund}.\\ 
Additional samples are generated with \textsc{Herwig++ 2.5.1} \cite{Herwigpp}, using the
CTEQ6.6 PDF \cite{pdf1} and the \textsc{UE7000} tune for 
the underlying event \cite{ue7k}. \textsc{Herwig++} uses angular-ordered parton showers, 
a cluster hadronisation scheme and its own underlying-event parameterisation given by \textsc{Jimmy} \cite{Jimmy}.\\ 
A different approach to simulate multi-jet final states is followed by \textsc{Alpgen} 
\cite{alpgen}. This approach is based on LO matrix-element calculations for $2\to n$ 
multi-parton final states, with $n\leq 6$, interfaced with \textsc{Herwig+Jimmy}
\cite{Herwig,Jimmy} to provide the parton shower, hadronisation and underlying-event 
models. \textsc{Alpgen} is known to provide a good description of the multi-jet 
final states as measured by ATLAS \cite{multijet}.
\section{Event selection and jet calibration}
\label{sec:evSel}
The data used in this analysis were recorded in 2011 at $\sqrt{s}=7~\TeV$ and  
collected using a single-jet trigger. It requires at least one jet,
reconstructed with the anti-$k_t$ algorithm \cite{jets} with radius parameter $R=0.4$ 
as implemented in \textsc{FastJet} \cite{fastjet}. The jet transverse energy, 
$\et = E \sin\theta$, is required to be greater than 135 $\GeV$ at the trigger level. This trigger is 
fully efficient at reconstructed transverse energies above
$240 \GeV$. Taking into account the prescale factor of this trigger, the data collected
correspond to an effective integrated luminosity of $\mathcal{L}_{\mathrm{eff}} = 158 \ \ipb$ \cite{lumi}.\\
Events are required to have at least one primary vertex, with five or more associated tracks 
with transverse momentum $\pt > 400 \MeV$. If there is more than one primary vertex, the vertex 
maximising $\sum \pt^2$ is chosen. MC simulated events are subject  
to a reweighting algorithm in order to match the average number of interactions
per bunch-crossing observed in the data.\\
In the analysis, jets are reconstructed with the same algorithm as used in the trigger, the anti-$k_t$ algorithm 
with radius parameter $R=0.4$. The input objects to the jet algorithm are topological clusters of energy deposits in the calorimeters
\cite{lampl}. The baseline calibration for these clusters corrects their energy using local hadronic 
calibration \cite{lc1,lc2}. The four-momentum of an uncalibrated jet is defined as the sum of
the four-momenta of its constituent clusters, which are considered massless. The resulting jets are massive.
However, the effect of this mass is marginal for jets in the kinematic range considered in this paper.\\
The jet calibration procedure includes energy corrections for multiple $pp$ interactions in the same or neighbouring bunch crossings, termed ``pileup''
in the following, as well as angular corrections to ensure that the jet originates from the primary vertex.
Effects due to energy losses in inactive material, shower
leakage, the magnetic field, as well as inefficiencies in energy clustering and jet
reconstruction, are taken into account. This is done using an MC-based correction, in bins of $\eta$ and $\pt$, derived from the relation of the reconstructed jet energy 
to the energy of the corresponding hadron-level jet, not including muons or non-interacting particles.
In a final step, an in situ calibration corrects for residual differences in the jet response between
the MC simulation and the data using momentum-balance techniques for
dijet, $\gamma+$jet, $\Zboson$+jet and multi-jet final states. This so-called jet energy scale (JES) \cite{JES} is 
subject to uncertainties including those affecting the energy of well-measured 
objects, like $\Zboson$ bosons and photons. The total JES uncertainty is given by a set of independent sources, 
correlated in $\pt$. The uncertainty in the $\pt$ of individual jets due to the 
JES increases from (1--4)$\%$ for $|\eta| < 1.8$, to $5\%$ for $1.8 <|\eta|<4.5$.\\
The selected events must have at least two jets with
transverse momentum $\pt > 50 \GeV$ and pseudorapidity 
$|\eta| < 2.5$. The two leading jets are further required 
to fulfil $p_{\mathrm{T}1} + p_{\mathrm{T}2} > 500 \GeV$. In addition, jets 
are required to satisfy quality criteria that reject beam-induced backgrounds \cite{cleaning}, 
as well as criteria for the fraction of the momentum of tracks within the jet which arise from the
primary interaction vertex. 
The number of selected events in data is $3.8\times 10^5$, 
with an average jet multiplicity $\langle N_{\mathrm{jet}} \rangle = 2.6$. The
resulting distribution for $(p_{\mathrm{T}1} + p_{\mathrm{T}2})/2$ extends up to $1.3 \TeV$
with an average value of $305 \GeV$.  
\section{Results at the detector level}
\label{sec:resultsDet}
\begin{figure}[h]
\centering
\includegraphics[width=8.cm,height=9.7cm]{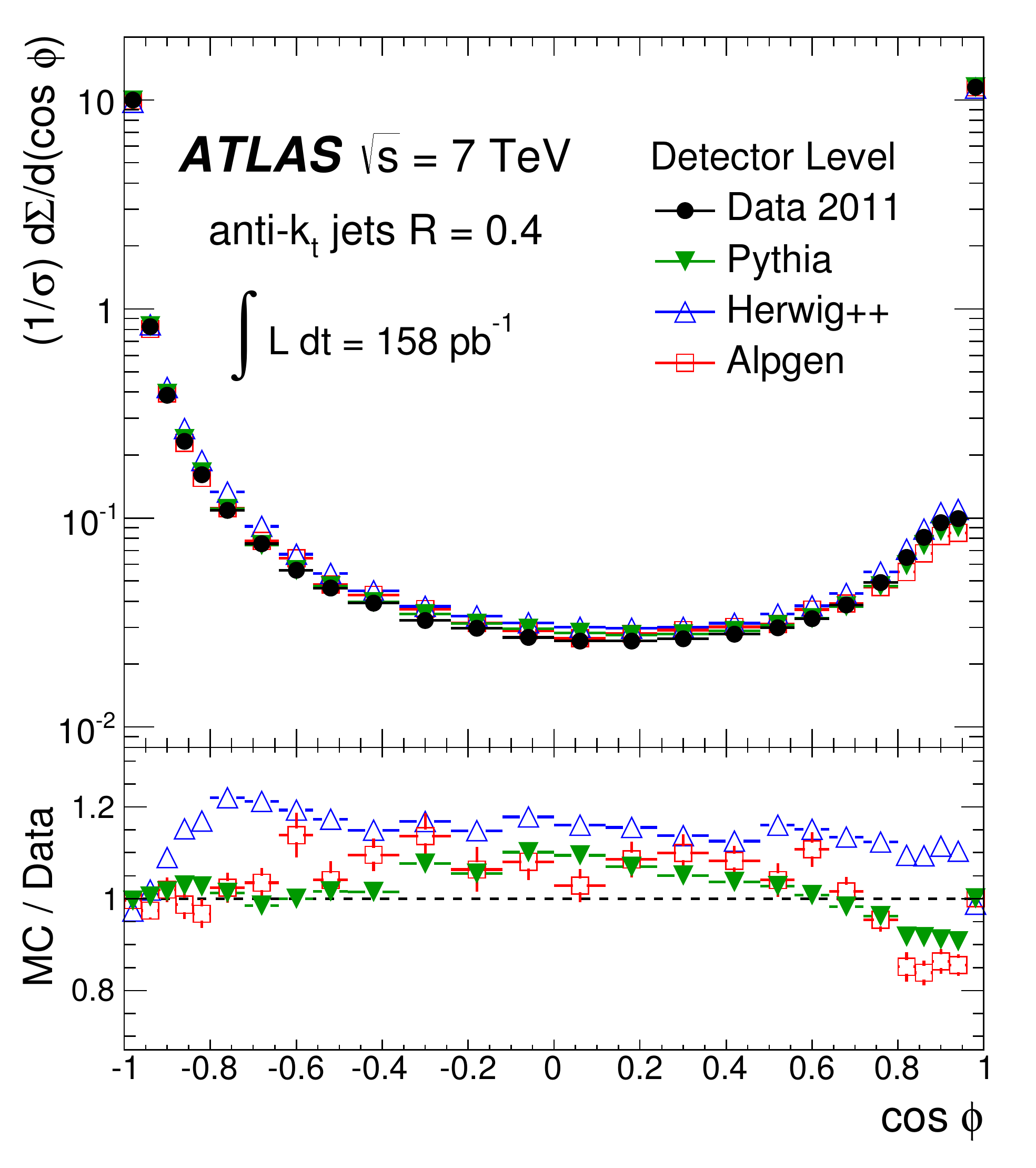}
\includegraphics[width=8.cm,height=9.7cm]{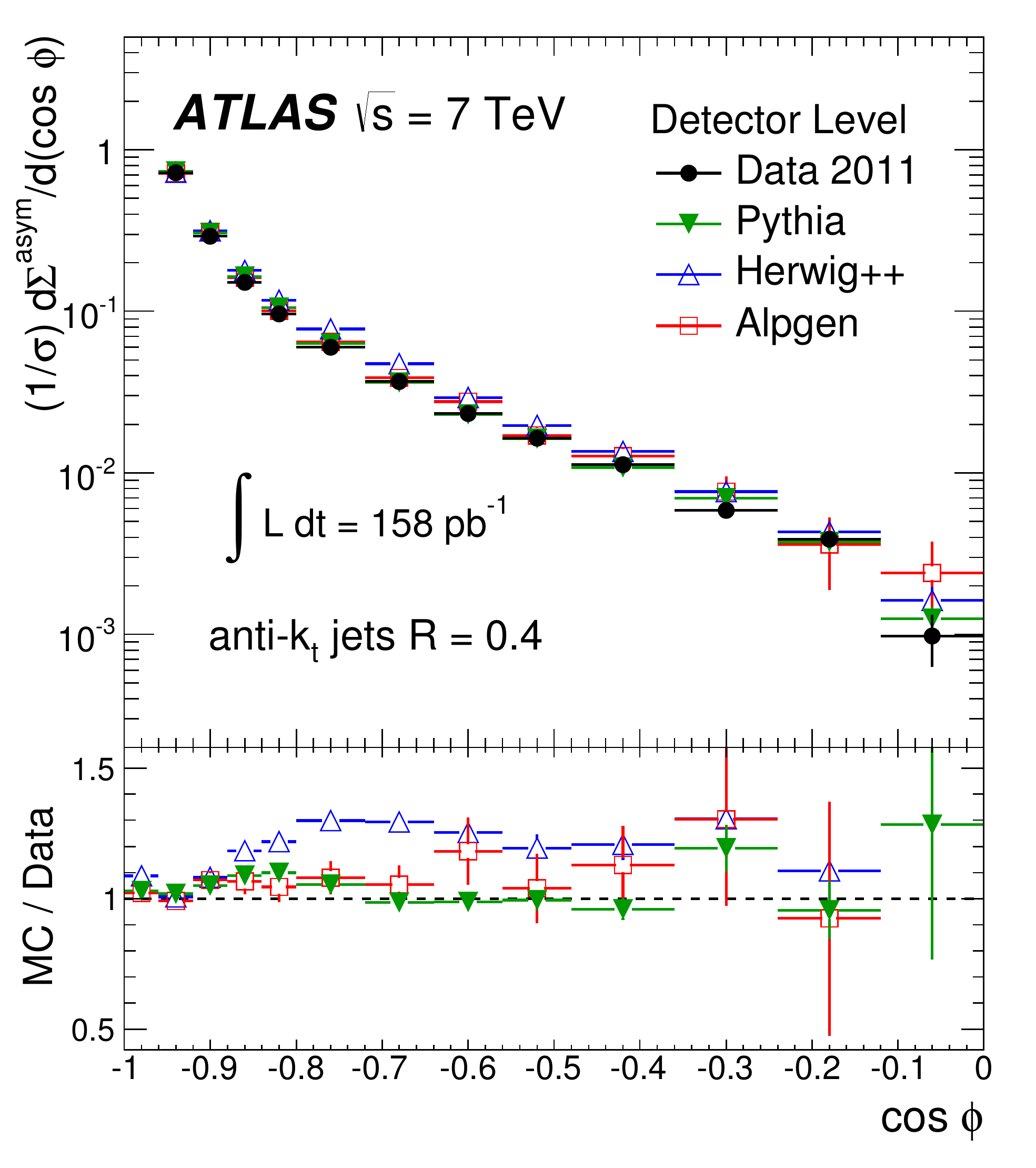}
\caption{The detector-level distributions for the transverse energy--energy correlation TEEC
(left) and its asymmetry ATEEC (right) along with comparisons to MC model
expectations. The uncertainties shown are statistical only. The first bin of the 
ATEEC distribution has a negative value and is therefore not included in the figure.}
\label{fig:recoEEC}
\end{figure}

The selected events are used to measure the TEEC and its associated asymmetry ATEEC, 
as defined in Equations (\ref{eq:defEEC}) and (\ref{eq:defAEEC}). The TEEC distribution for a sample of $N$ events is 
obtained by calculating the cosines of the angles in the transverse plane between 
all possible pairs of jets in each event. Every pair $(i,j)$ 
represents an entry in the distribution, which is then weighted with 
the normalised product of the transverse energies. The 
weights $w_{ij}$ are defined as 
\begin{equation}
\hspace{-0.5cm} w_{ij} = \xti\xtj = \frac{E_{\mathrm{T}i}E_{\mathrm{T}j}}{\left(\sum_k E_{\mathrm{T}k}\right)^2},
\label{eq:etWeights}
\end{equation}
such that for a given event their sum is always unity, as the self correlations 
$i=j$ are also taken into account. The resulting distribution is then divided by the number of events, which
normalises it to unit area. This weighting procedure reduces the sensitivity to the jet energy scale 
and resolution.\\
Figure \ref{fig:recoEEC} shows the TEEC and ATEEC distributions along with comparisons to 
detector-level \textsc{Pythia}, \textsc{Herwig} and \textsc{Alpgen} expectations. The 
TEEC exhibits peaks at $\cos \phi=1$ (self correlations) and near $\cos \phi=-1$, with 
a rather flat central region around $\cos \phi = 0$. These features are similar 
to those observed in $e^+e^-$ annihilation, as described in Ref.~\cite{DELPHI}. The 
central region is expected to be dominated by hard radiation processes while multiple 
soft radiation is expected to be important in the $\cos\phi \simeq \pm 1$ regions.\\
The description of the TEEC is good in the back-to-back region $\cos\phi \simeq -1$ for both \textsc{Pythia 6} and \textsc{Alpgen}. Differences
up to 10\% are observed in the central part, while the region of small angles shows differences as large as
about 15\%. The description by \textsc{Herwig++} is poorer. The 
ATEEC exhibits a steep fall-off, which is reproduced by both \textsc{Pythia 6} and 
\textsc{Alpgen}. \textsc{Herwig++} shows some discrepancies as large as 30\%.
\section{Correction to particle level}
\label{sec:unf}
\begin{figure}[h]
\centering
\includegraphics[width=8.cm,height=7.cm]{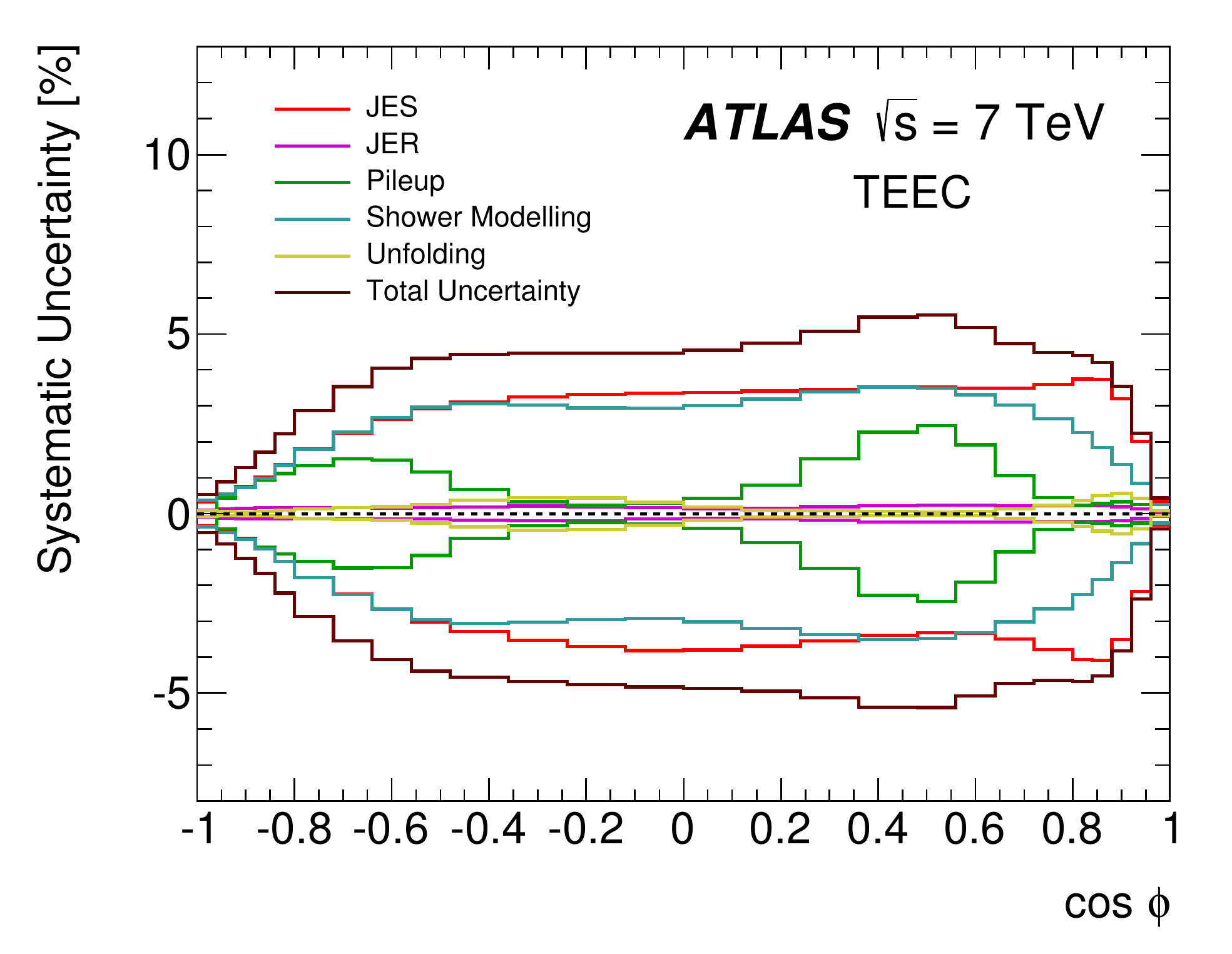}
\includegraphics[width=8.cm,height=7.cm]{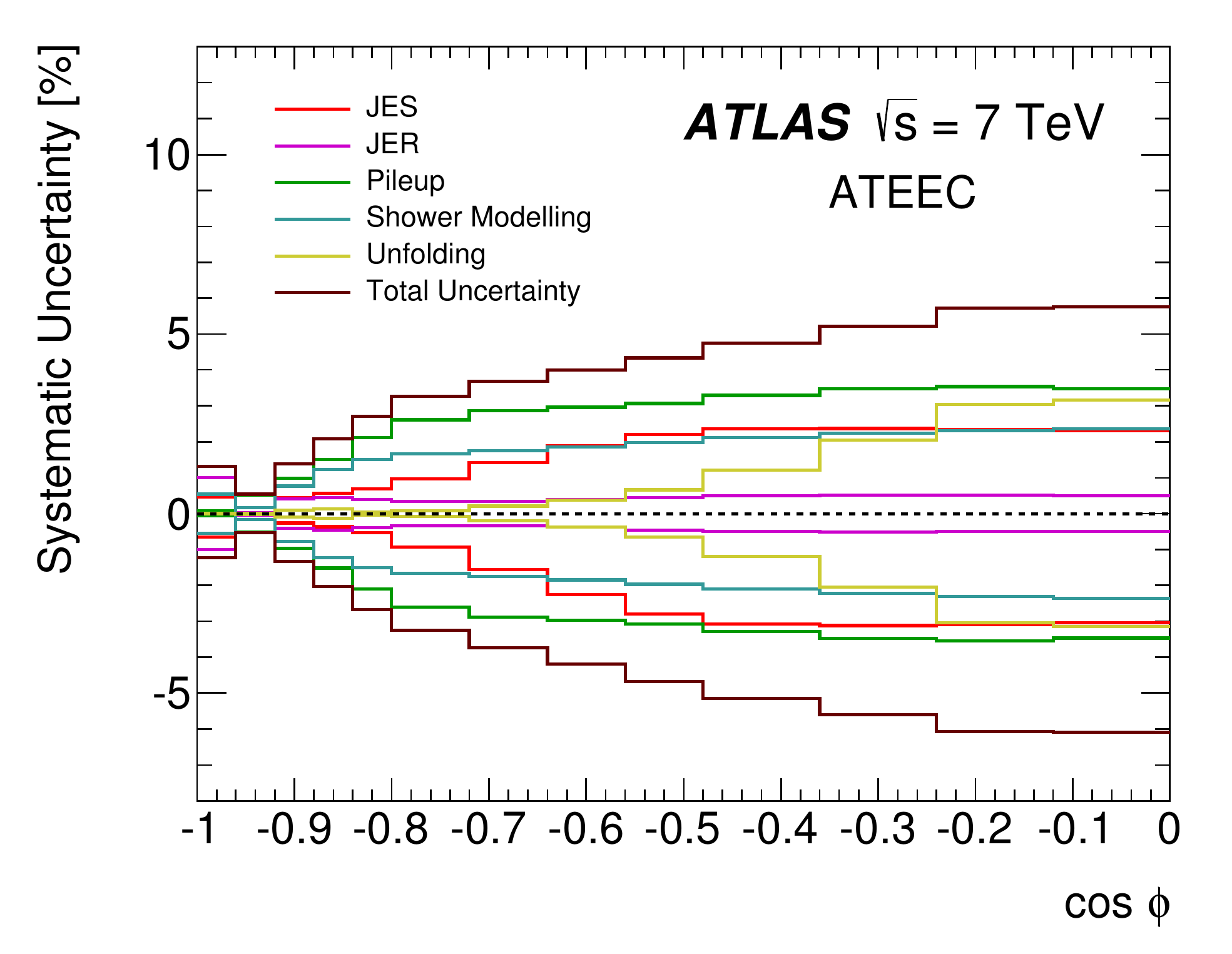}
\caption{Relative systematic uncertainties for the TEEC (left) and the ATEEC (right) as a function of $\cos\phi$.}
\label{fig:sysUnc}
\end{figure}

The data are corrected to the particle level in order to take into account 
detector efficiencies and resolutions. This allows a direct comparison with theoretical calculations, as 
well as with measurements of other experiments. \\
Particle-level jets are reconstructed in MC events using all particles with average lifetime 
$\tau > 10^{-11} \mathrm{s}$, including muons and neutrinos. The kinematic 
selection criteria are the same as for the detector-level distribution. The 
unfolding relies on a bin-by-bin correction given by the ratios 
of the particle-level to detector-level distributions in the \textsc{Pythia AUET2B} sample, 
which is then applied to the detector-level distributions in data.
To check the effect of bin migrations on the unfolding procedure, an iterative Bayesian method \cite{bayes}
as implemented in \textsc{RooUnfold} \cite{roounfold} is also used. The convergence criteria is fulfilled when the linear sum over all bins 
of the absolute relative differences from one iteration to the next drops below $10^{-2}$. The method converges after five iterations.
The differences between the two approaches are negligible, compared to the statistical uncertainties,
in the full range of $\cos\phi$. This is expected due to the high azimuthal resolution of the jet axis, which is 10 mrad.\\
The following experimental sources of uncertainty are considered for this measurement:
\begin{itemize}
\item{\bf{Jet Energy Scale:}} The uncertainty due to the jet energy scale (JES) 
\cite{JES} is calculated using MC techniques by varying each jet
energy and momentum by one standard deviation for each of the 63 independent sources
of the JES uncertainty, and propagated to the TEEC.
These uncertainties depend on the jet transverse momentum and pseudorapidity. The total uncertainty due to the JES is calculated as the sum in quadrature
of all independent uncertainties. In order to investigate the effect of possible
correlations between JES sources in the analysis, two alternative scenarios 
with weaker and stronger correlations have been considered \cite{JES}. The impact of the 
change of correlation configurations, as well as of the number of JES independent 
sources, on the value of $\alpha_{\mathrm{s}}(m_\Zboson)$ and its experimental error is found 
to be negligible.\\
The values of the JES uncertainty are typically asymmetric for both the TEEC and ATEEC distributions,
although the values for this asymmetry are small. Thus, the positive and negative parts of the uncertainty are independently summed in quadrature.\\
The TEEC distribution has a total uncertainty of up to 3.5\% from the JES sources,  
the largest contributions being due to close-by jets and to the different response to jets
initiated by gluons or quarks. This is the dominant experimental systematic uncertainty in the analysis.
\item{\bf{Jet Energy Resolution:}} The uncertainty in the jet energy resolution  
\cite{JER} is propagated to the TEEC by smearing each jet transverse momentum by 
a $\pt$- and $\eta$-dependent factor accounting for the resolution uncertainty. The size of this 
uncertainty is below 1\% for both the TEEC and the ATEEC distributions.
\item{\bf{Pileup:}} The pileup uncertainty is estimated by comparing the ratio 
of the detector-level TEEC and ATEEC distributions obtained in samples with 
reduced ($\mu<6$) and enhanced pileup activity ($\mu>6$). Here $\mu$ is the
average number of interactions per bunch crossing \cite{lumi}. These ratios are formed in both
data and MC simulation and the difference is assigned as the pileup systematic uncertainty,
which is as large as 2\% (4\%) for the TEEC (ATEEC). The size of this dedicated
estimate is larger than what is predicted by the sum of the two sources of
uncertainty due to pileup included in the JES uncertainty. The envelope of the two different estimates is used.
\item{\bf{Parton shower modelling:}} To estimate the uncertainty due to the parton 
shower modelling, the data unfolded with \textsc{Pythia 6} 
and \textsc{Herwig++} are compared. The parton shower and
hadronisation models in the two generators are different, as is the implementation 
of UE effects. The size of this uncertainty is as large as 3.5\% (2.5\%) for the TEEC (ATEEC).
\item{\bf{Unfolding:}} To estimate the uncertainty associated with the unfolding
 procedure, a data-driven method is used to test its stability.
This method relies on the reweighting of the particle-level projection of the unfolding transfer matrix so that 
the agreement between the detector-level projection and the data is enhanced. This 
modified detector-level distribution is then unfolded using the correction 
factors described above. The difference between the modified 
particle-level distribution and the nominal one is then taken as the 
uncertainty. This uncertainty is smaller than 0.5\% for the full $\cos\phi$ range.
\end{itemize}
Other possible sources of uncertainty are also studied, such as the jet 
angular resolution and jet quality selection procedure. They
are found to be at the per mille level, much smaller than the statistical 
uncertainty on the corrected data, and are therefore neglected.
To reduce the effect of statistical fluctuations, all the independent 
systematic uncertainties discussed here are smoothed separately.
Figure \ref{fig:sysUnc} shows the breakdown of the systematic uncertainties for 
both the TEEC and the ATEEC, together with the total, obtained as the sum 
in quadrature of every independent source discussed above.

\begin{figure}[h]
\centering
\includegraphics[width=8.cm,height=9.7cm]{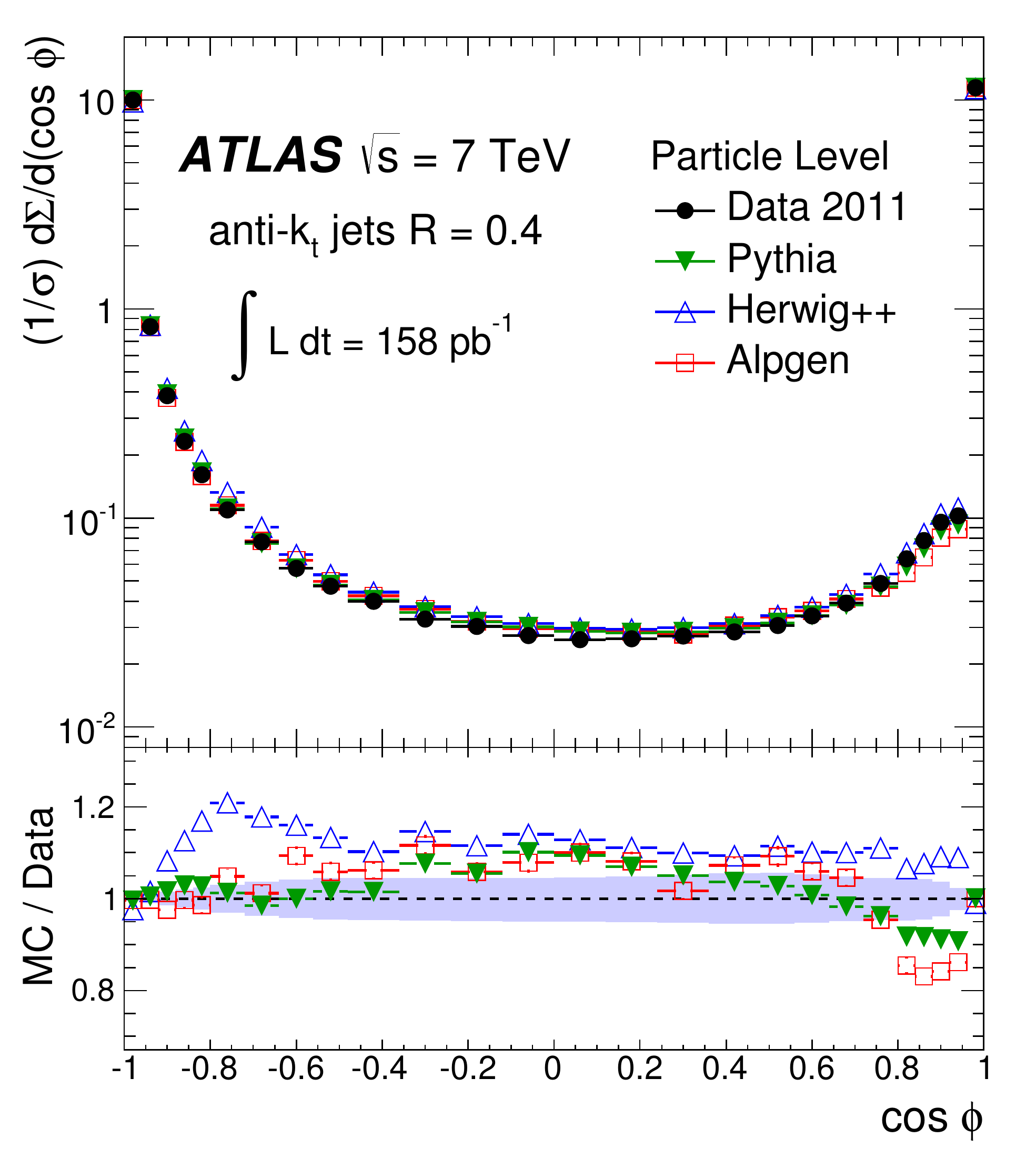}
\includegraphics[width=8.cm,height=9.7cm]{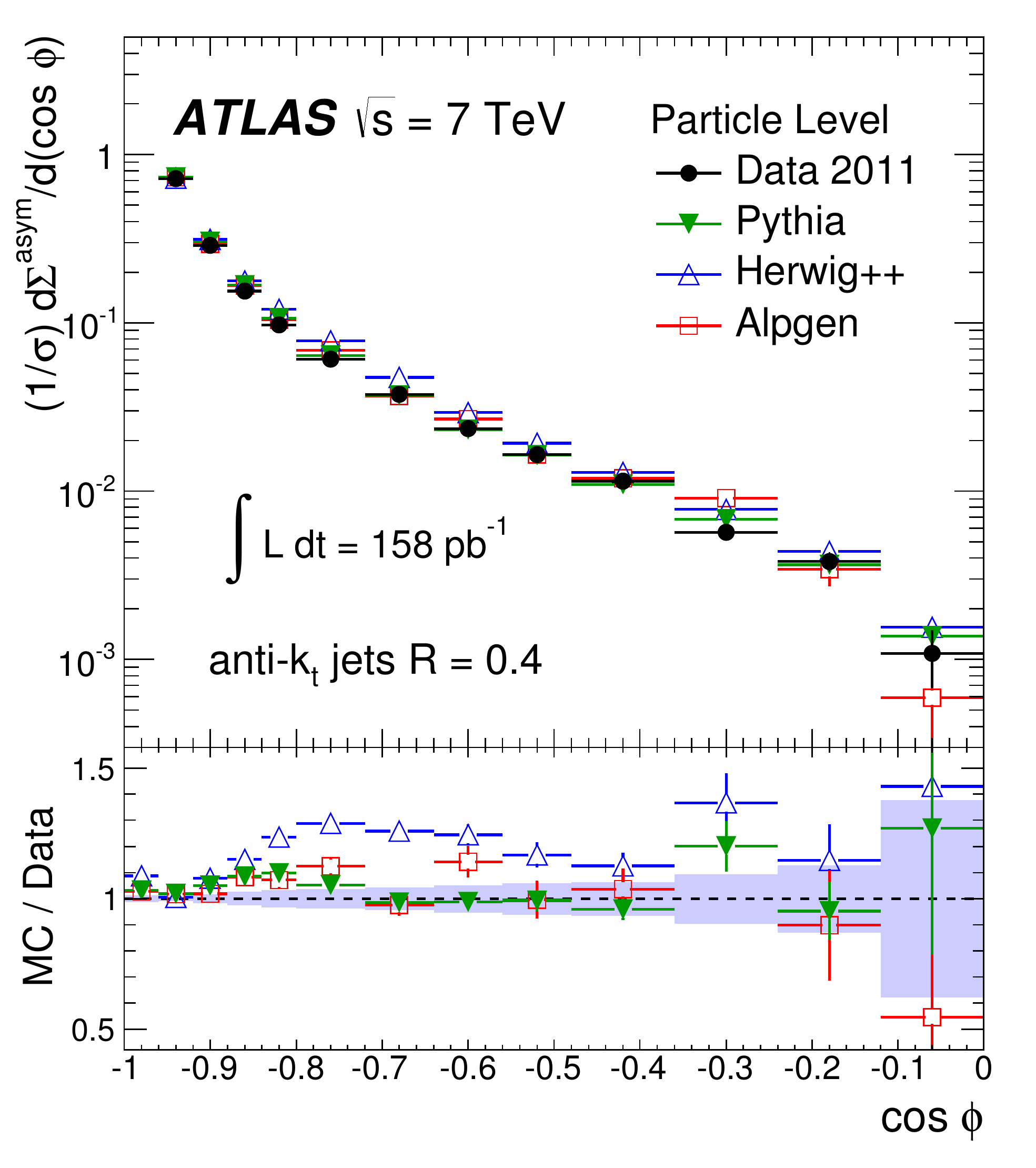}
\caption{The unfolded distributions for transverse energy--energy correlation (left)
and its asymmetry (right) along with comparisons to MC expectations. The
statistical uncertainties are shown with error bars, while the total experimental
uncertainties are shown in a shaded band.}
\label{fig:hadLevel}
\end{figure}

The TEEC and ATEEC distributions, once corrected for detector effects, are shown in Figure \ref{fig:hadLevel},
together with their total uncertainties, while numerical values 
are given in Tables \ref{tabTEEC} and \ref{tabATEEC}.

As already seen in the detector-level distributions, \textsc{Pythia 6} and \textsc{Alpgen} give a fair description of the data
both for the TEEC and ATEEC. The back-to-back region $\cos \phi \sim -1$
is well described, while small discrepancies, at the level of $10\%$, 
are observed in the central region of the TEEC and for large $\cos \phi$ values.
The description by \textsc{Herwig++} is poorer.\\
The shape of the ATEEC is very  similar to that observed at $e^+e^-$ colliders,
see Refs. \cite{PLUTO,MARKII,MARKJ,CELLO,JADE,MAC,TASSO,TOPAZ,DELPHI,OPAL,ALEPH,L3,SLD}, 
and well reproduced by \textsc{Pythia 6} and \textsc{Alpgen}.

\begin{table}[h]
\begin{center}
\begin{tabular}{cccccccc}
\hline
$\cos \phi$ & TEEC  & Stat. & JES & JER & Shower & Pileup & Unfolding\\
\hline
\TA\B $( -1.00 , -0.96 )$ & 10.008 & 0.008 & $^{+ 0.033 }_{- 0.034 }$ & 0.009 & 0.037 & 0.008 & 0.008 \\
\TA\B $( -0.96 , -0.92 )$ & 0.8218 & 0.0047 & $^{+ 0.0044 }_{- 0.0040 }$ & 0.0011 & 0.0044 & 0.0036 & 0.0005 \\
\TA\B $( -0.92 , -0.88 )$ & 0.3848 & 0.0029 & $^{+ 0.0029 }_{- 0.0026 }$ & 0.0006 & 0.0028 & 0.0028 & 0.0002 \\
\TA\B $( -0.88 , -0.84 )$ & 0.2324 & 0.0022 & $^{+ 0.0024 }_{- 0.0022 }$ & 0.0004 & 0.0023 & 0.0022 & 0.0001 \\
\TA\B $( -0.84 , -0.80 )$ & 0.1612 & 0.0017 & $^{+ 0.0022 }_{- 0.0022 }$ & 0.0003 & 0.0022 & 0.0018 & 0.0002 \\
\TA\B $( -0.80 , -0.72 )$ & 0.1095 & 0.0009 & $^{+ 0.0020 }_{- 0.0020 }$ & 0.0002 & 0.0020 & 0.0015 & 0.0002 \\
\TA\B $( -0.72 , -0.64 )$ & 0.0767 & 0.0008 & $^{+ 0.0017 }_{- 0.0017 }$ & 0.0001 & 0.0017 & 0.0012 & 0.0001 \\
\TA\B $( -0.64 , -0.56 )$ & 0.0574 & 0.0006 & $^{+ 0.0015 }_{- 0.0015 }$ & 0.0001 & 0.0015 & 0.0009 & 0.0001 \\
\TA\B $( -0.56 , -0.48 )$ & 0.0472 & 0.0005 & $^{+ 0.0014 }_{- 0.0014 }$ & 0.0001 & 0.0014 & 0.0005 & 0.0001 \\
\TA\B $( -0.48 , -0.36 )$ & 0.0400 & 0.0004 & $^{+ 0.0012 }_{- 0.0013 }$ & 0.0001 & 0.0012 & 0.0003 & 0.0001 \\
\TA\B $( -0.36 , -0.24 )$ & 0.0329 & 0.0004 & $^{+ 0.0011 }_{- 0.0012 }$ & 0.0001 & 0.0010 & 0.0001 & 0.0001 \\
\TA\B $( -0.24 , -0.12 )$ & 0.0302 & 0.0003 & $^{+ 0.0010 }_{- 0.0011 }$ & 0.0001 & 0.0009 & 0.0001 & 0.0001 \\
\TA\B $( -0.12 , 0.00 )$ & 0.0273 & 0.0003 & $^{+ 0.0009 }_{- 0.0010 }$ & -- & 0.0008 & 0.0001 & 0.0001 \\
\TA\B $( 0.00 , 0.12 )$ & 0.0262 & 0.0003 & $^{+ 0.0009 }_{- 0.0010 }$ & -- & 0.0008 & 0.0001 & -- \\
\TA\B $( 0.12 , 0.24 )$ & 0.0264 & 0.0003 & $^{+ 0.0009 }_{- 0.0010 }$ & -- & 0.0008 & 0.0002 & -- \\
\TA\B $( 0.24 , 0.36 )$ & 0.0272 & 0.0003 & $^{+ 0.0009 }_{- 0.0010 }$ & 0.0001 & 0.0009 & 0.0004 & -- \\
\TA\B $( 0.36 , 0.48 )$ & 0.0286 & 0.0003 & $^{+ 0.0010 }_{- 0.0010 }$ & 0.0001 & 0.0010 & 0.0006 & -- \\
\TA\B $( 0.48 , 0.56 )$ & 0.0306 & 0.0004 & $^{+ 0.0011 }_{- 0.0010 }$ & 0.0001 & 0.0011 & 0.0008 & -- \\
\TA\B $( 0.56 , 0.64 )$ & 0.0340 & 0.0004 & $^{+ 0.0012 }_{- 0.0011 }$ & 0.0001 & 0.0011 & 0.0006 & -- \\
\TA\B $( 0.64 , 0.72 )$ & 0.0391 & 0.0004 & $^{+ 0.0014 }_{- 0.0014 }$ & 0.0001 & 0.0012 & 0.0004 & 0.0001 \\
\TA\B $( 0.72 , 0.80 )$ & 0.0487 & 0.0004 & $^{+ 0.0017 }_{- 0.0018 }$ & 0.0001 & 0.0013 & 0.0002 & 0.0001 \\
\TA\B $( 0.80 , 0.84 )$ & 0.0639 & 0.0007 & $^{+ 0.0024 }_{- 0.0026 }$ & 0.0001 & 0.0014 & 0.0002 & 0.0002 \\
\TA\B $( 0.84 , 0.88 )$ & 0.0780 & 0.0008 & $^{+ 0.0029 }_{- 0.0032 }$ & 0.0002 & 0.0014 & 0.0002 & 0.0004 \\
\TA\B $( 0.88 , 0.92 )$ & 0.0955 & 0.0009 & $^{+ 0.0031 }_{- 0.0033 }$ & 0.0002 & 0.0013 & 0.0003 & 0.0005 \\
\TA\B $( 0.92 , 0.96 )$ & 0.1025 & 0.0009 & $^{+ 0.0021 }_{- 0.0022 }$ & 0.0001 & 0.0009 & 0.0003 & 0.0004 \\
\TA\B $( 0.96 , 1.00 )$ & 11.448 & 0.003 & $^{+ 0.039 }_{- 0.036 }$ & 0.006 & 0.030 & 0.008 & 0.008 \\
\hline
\end{tabular}
\end{center}
\caption{Values of the transverse energy--energy correlation function (TEEC). The statistical and systematic uncertainties,
due to Jet Energy Scale and Resolution (JES and JER), shower modelling as well as pileup and unfolding, are shown in the subsequent columns. Uncertainties marked with a dash (--) are smaller than 0.00005.}
\label{tabTEEC}
\end{table}

\begin{table}[h]
\begin{center}

\begin{tabular}{ccccccccc}
\hline
$\cos \phi$ & ATEEC  & Stat. & JES & JER & Shower & Pileup & Unfolding\\
\hline
\TA\B $( -1.00 , -0.96 )$ & $-$1.4406 & 0.0083 & $^{+ 0.0094 }_{- 0.0066 }$ & 0.0144 & 0.0078 & 0.0010 & 0.0001 \\
\TA\B $( -0.96 , -0.92 )$ & 0.7193 & 0.0048 & $^{+ 0.0002 }_{- 0.0000 }$ & 0.0002 & 0.0012 & 0.0037 & 0.0001 \\
\TA\B $( -0.92 , -0.88 )$ & 0.2893 & 0.0030 & $^{+ 0.0012 }_{- 0.0008 }$ & 0.0012 & 0.0022 & 0.0028 & 0.0003 \\
\TA\B $( -0.88 , -0.84 )$ & 0.1544 & 0.0023 & $^{+ 0.0009 }_{- 0.0006 }$ & 0.0007 & 0.0019 & 0.0023 & 0.0002 \\
\TA\B $( -0.84 , -0.80 )$ & 0.0973 & 0.0019 & $^{+ 0.0007 }_{- 0.0005 }$ & 0.0004 & 0.0015 & 0.0020 & -- \\
\TA\B $( -0.80 , -0.72 )$ & 0.0608 & 0.0010 & $^{+ 0.0006 }_{- 0.0006 }$ & 0.0002 & 0.0010 & 0.0016 & -- \\
\TA\B $( -0.72 , -0.64 )$ & 0.0376 & 0.0009 & $^{+ 0.0005 }_{- 0.0006 }$ & 0.0001 & 0.0007 & 0.0011 & 0.0001 \\
\TA\B $( -0.64 , -0.56 )$ & 0.0235 & 0.0007 & $^{+ 0.0004 }_{- 0.0005 }$ & 0.0001 & 0.0004 & 0.0007 & 0.0001 \\
\TA\B $( -0.56 , -0.48 )$ & 0.0165 & 0.0007 & $^{+ 0.0004 }_{- 0.0005 }$ & 0.0001 & 0.0003 & 0.0005 & 0.0001 \\
\TA\B $( -0.48 , -0.36 )$ & 0.0115 & 0.0005 & $^{+ 0.0003 }_{- 0.0004 }$ & 0.0001 & 0.0002 & 0.0004 & 0.0001 \\
\TA\B $( -0.36 , -0.24 )$ & 0.0057 & 0.0004 & $^{+ 0.0001 }_{- 0.0002 }$ & -- & 0.0001 & 0.0002 & 0.0001 \\
\TA\B $( -0.24 , -0.12 )$ & 0.0038 & 0.0004 & $^{+ 0.0001 }_{- 0.0001 }$ & -- & 0.0001 & 0.0001 & 0.0001 \\
\TA\B $( -0.12 , 0.00 )$ & 0.0011 & 0.0004 & -- & -- & -- & -- & -- \\
\hline
\end{tabular}
\end{center}
\caption{Values of the asymmetry on the transverse energy--energy correlation 
function (ATEEC). The statistical and systematic uncertainties, 
due to Jet Energy Scale and Resolution (JES and JER), shower modelling as well
as pileup and unfolding, are shown in the subsequent columns. Uncertainties marked with a dash (--) are smaller than 0.00005.} 
\label{tabATEEC}
\end{table}

\section{Theoretical predictions and uncertainties}
\label{sec:theoPred}
In perturbative QCD (pQCD), according to the factorisation theorem \cite{factorTheorem}, final-state 
observables can be expressed as a convolution of the partonic cross-sections, $\hat{\sigma}$, 
with the parton distribution functions. Thus, in this particular case, the TEEC 
distribution to leading order in the strong coupling constant,
can be expressed as the three-jet, energy-weighted, differential cross-section in $\cos\phi$, 
normalised to the integrated two-jet cross-section. This can be schematically expressed as
\begin{equation}
\frac{1}{\sigma}\frac{\mathrm{d}\Sigma}{\mathrm{d}(\cos\phi)} =
\frac{\Sigma_{a_i,b_i} f_{a_1}(x_1) f_{a_2}(x_2) \otimes \hat{\Sigma}^{a_1 a_2 \to b_1 b_2 b_3}}
     {\Sigma_{a_i,b_i} f_{a_1}(x_1) f_{a_2}(x_2) \otimes \hat{\sigma}^{a_1 a_2 \to b_1 b_2 } },
\label{eq:TEEC-QCD}
\end{equation}
where $\hat{\Sigma}^{a_1 a_2 \to b_1 b_2 b_3}$ is the transverse energy--energy 
weighted partonic cross-section, $x_i$ ($ i=1,2$) are the fractional longitudinal 
momenta of the initial-state partons, $f_{a_1}(x_1)$ and $f_{a_2}(x_2)$ are the PDF, 
and $\otimes$ denotes a convolution over the appropriate variables. The denominator of Eq. (\ref{eq:TEEC-QCD})
is the integrated dijet cross-section used to normalise the TEEC.\\
\\
The pQCD NLO calculations of the TEEC and ATEEC distributions are performed 
using \textsc{NLOJet++} \cite{Nagy1,Nagy2} interfaced with the \textsc{MSTW 2008} \cite{mstw}, 
\textsc{CT10} \cite{ct10}, \textsc{NNPDF 2.3} \cite{nnpdf} and \textsc{HERAPDF 1.5} 
\cite{herapdf} parton distribution functions at NNLO. Typically, $\mathcal{O}(10^{10})$ events 
are generated for these calculations.  This involves the calculation of the 
$2\to 3$ partonic subprocesses at NLO accuracy and of the $2\to 4$ partonic subprocesses at tree level.
In order to avoid the double collinear singularities appearing in the latter \cite{Ali2013}, 
the angular range is restricted to $|\cos\phi| < 0.92$.\\
\\
The renormalisation and factorisation scales, inherent 
in any pQCD calculation, are usually taken to reflect the typical transverse momentum 
of the process under investigation. For the TEEC and ATEEC calculations, they are taken to be
\begin{equation}
\mu_{\mathrm{R}} = \mu_{\mathrm{F}} = \frac{p_{\mathrm{T}1}+p_{\mathrm{T}2}}{2},
\end{equation}
where $p_{\mathrm{T}1}$ and $p_{\mathrm{T}2}$ are the transverse momenta of 
the two leading jets. This is also the choice in Ref.~\cite{cmsAs}.
The value of the strong coupling constant at a given scale is connected to $\alpha_{\mathrm{s}}(m_\Zboson)$
using the two-loop beta function \cite{Nagy1,Nagy2}.\\
\\
The NLO theoretical predictions are subsequently corrected for non-perturbative 
effects such as hadronisation and the underlying event. This correction is calculated 
using the leading-logarithm parton shower generators \textsc{Pythia 6} and 
\textsc{Herwig++} interfaced with different tunes. The full MC generator particle-level
predictions with these effects switched on are compared with the parton-level 
predictions before hadronisation and without UE effects. From this comparison
a bin-by-bin correction factor is calculated as the ratio of the two predictions, which is then used to correct the 
\textsc{NLOJet++} output. They are found to deviate from unity by about 1\% for both 
\textsc{Pythia 6} and \textsc{Herwig++} for most of the $|\cos\phi|<0.92$ range.\\
\\
Three main theoretical uncertainties are considered for the analysis: those 
corresponding to the renormalisation and factorisation scale variations, those 
corresponding to the PDF, and those on the non-perturbative corrections.
\begin{itemize}
\item{Scale uncertainty:} The ambiguity in the choice of the renormalisation and factorisation 
scales gives rise to a scale uncertainty. To estimate it, the scales $\mu_{\mathrm{R}}$ and $\mu_{\mathrm{F}}$ 
are varied by a factor of two up and down, with the additional requirement that 
$0.5 \leq \mu_{\mathrm{R}} / \mu_{\mathrm{F}} \leq 2$. From all those variations, the largest uncertainty
is obtained when both $\mu_{\mathrm{R}}$ and $\mu_{\mathrm{F}}$ are varied simultaneously by the same factor
from the nominal scale. These two combinations are used to define the envelope
of the scale uncertainty for both the TEEC and ATEEC.
The size of the scale uncertainty is highly asymmetric and is at most about 8\% for 
the TEEC distribution, and somewhat smaller for the ATEEC.
\item{PDF uncertainty:} The \textsc{CT10} parton distribution functions provide 50 
variations for the 25 fitted parameters at the 90\% confidence level. Each 
of the 25 parameters are varied up and down following the \textsc{CT10} 
recommendations in Ref.~\cite{ct10}, and are combined for each bin of the TEEC 
and ATEEC distributions following the prescription given in Ref.~\cite{primer}.
The size of the PDF uncertainty, once scaled at 68\% confidence level, is about $1.5\%$. A 
similar procedure is used for the \textsc{MSTW2008}, 
\textsc{NNPDF 2.3} and \textsc{HERAPDF 1.5} parton distribution functions. 
\item{Uncertainties in the non-perturbative corrections:} The 
non-perturbative corrections (NPC) are
calculated using \textsc{Pythia 6} interfaced to the \textsc{AUET2B} 
and \textsc{AMBT2B} tunes \cite{atlasTunes,atlasTunes2}, as well as 
\textsc{Herwig++} with the \textsc{UE7000} tune \cite{ue7k}. Moreover, 
\textsc{Pythia 8} interfaced to the \textsc{4C} and \textsc{AU2} tunes is also used. 
An uncertainty is derived by considering, on a bin-by-bin basis, the maximum 
difference between the nominal \textsc{Pythia AUET2B} and any other tune. Its 
size is below 1\% for most of the angular range considered.
\end{itemize}

\section{Determination of the strong coupling $\alpha_{\mathrm{s}}(m_\Zboson)$}
\label{sec:alphaS}
The evaluation of $\alpha_{\mathrm{s}}(m_\Zboson)$ is made by minimising a $\chi^2$
function taking into account correlations between the systematic uncertainties
using nuisance parameters $\lambda_k$, one for each source
of uncertainty. These nuisance parameters are normalised to zero mean and unit variance.
The minimum of the $\chi^2$ function is found in a 66-dimensional 
space, one dimension corresponding to $\alpha_{\mathrm{s}}(m_\Zboson)$ and the rest to the 
nuisance parameters associated with the experimental errors. The function to be 
minimised is defined as
\begin{equation}
\chi^2(\alpha_{\mathrm{s}},\vec{\lambda}) = \sum_{i}\frac{(x_i-F_i(\alpha_{\mathrm{s}},\vec{\lambda}))^2}{\Delta x_i^2+\Delta\tau_i ^2} + 
\sum_{k}\lambda_k^2,
\label{eq:chi2a}
\end{equation}
where the \textsc{NLOJet++} predictions are varied according to
\begin{equation}
F_{i}(\alpha_{\mathrm{s}},\vec{\lambda}) = \psi_{i}(\alpha_{\mathrm{s}})\left(1+\sum_k\lambda_k\sigma_k^{(i)}\right).
\label{eq:chi2b}
\end{equation}
In these expressions, $x_i$ corresponds to the data points in each distribution 
(TEEC or ATEEC), and $\Delta x_i$ are their statistical uncertainties. $\Delta\tau_i$ 
are the statistical errors on the \textsc{NLOJet++} predictions, while 
$\sigma_k^{(i)}$ correspond to the $k$-th source of experimental uncertainty in 
the bin $i$.\\
The functions $\psi_i(\alpha_{\mathrm{s}})$ are analytical expressions parameterising 
the dependence of each observable (TEEC or ATEEC) on the strong coupling 
constant. They are obtained by fitting the predictions for each bin as a 
function of $\alpha_{\mathrm{s}}(m_\Zboson)$. This function is chosen to be a parabola, 
as the theoretical predictions account 
for terms quadratic in $\alpha_{\mathrm{s}}$. The quality of the fit to the 
NLO theoretical predictions is found to be 
excellent for each bin of the TEEC and ATEEC. The uncertainties from these fits are negligible.

\begin{figure}[h]
\centering
\includegraphics[width=8.cm,height=9.5cm]{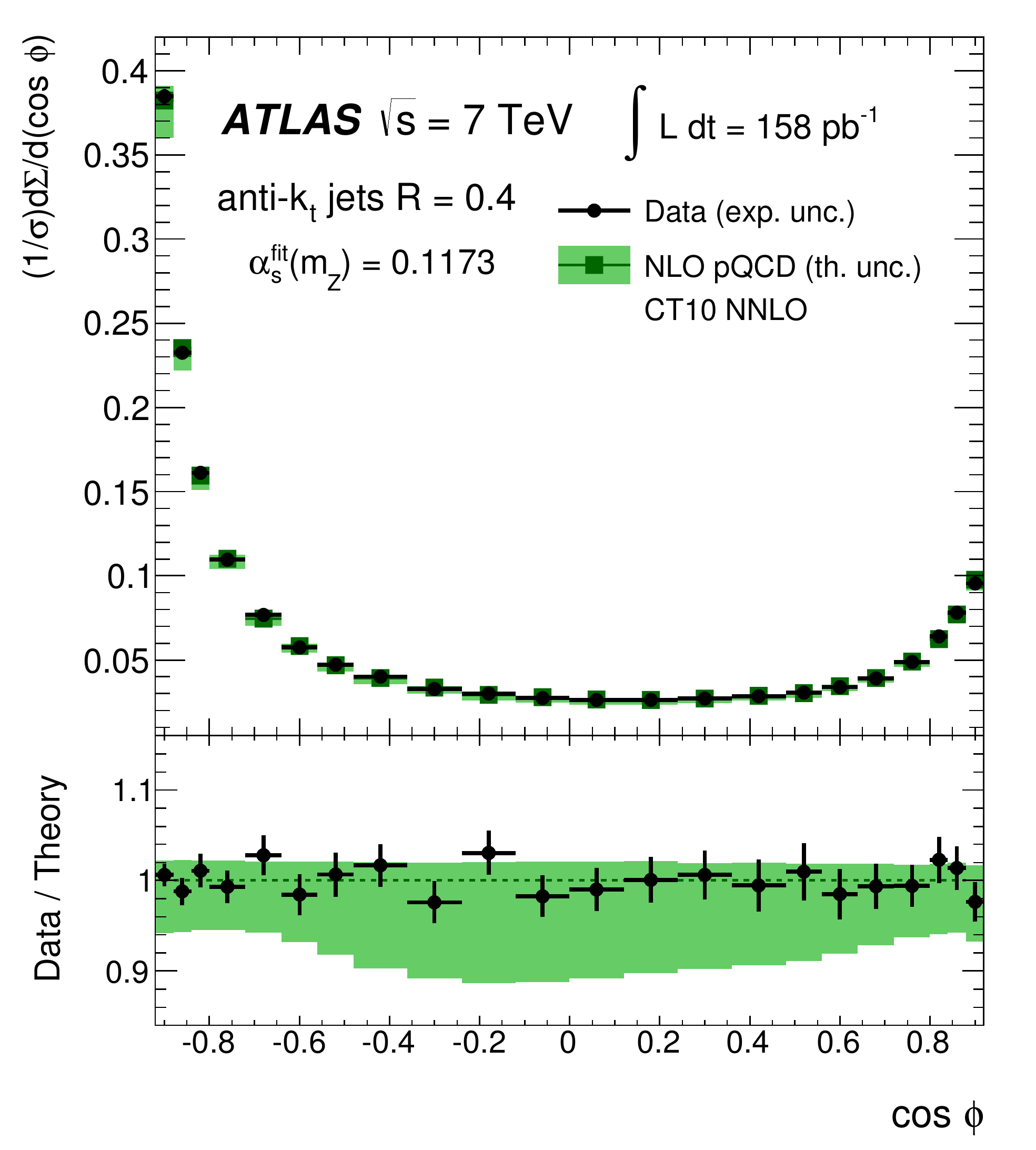}
\includegraphics[width=8.cm,height=9.5cm]{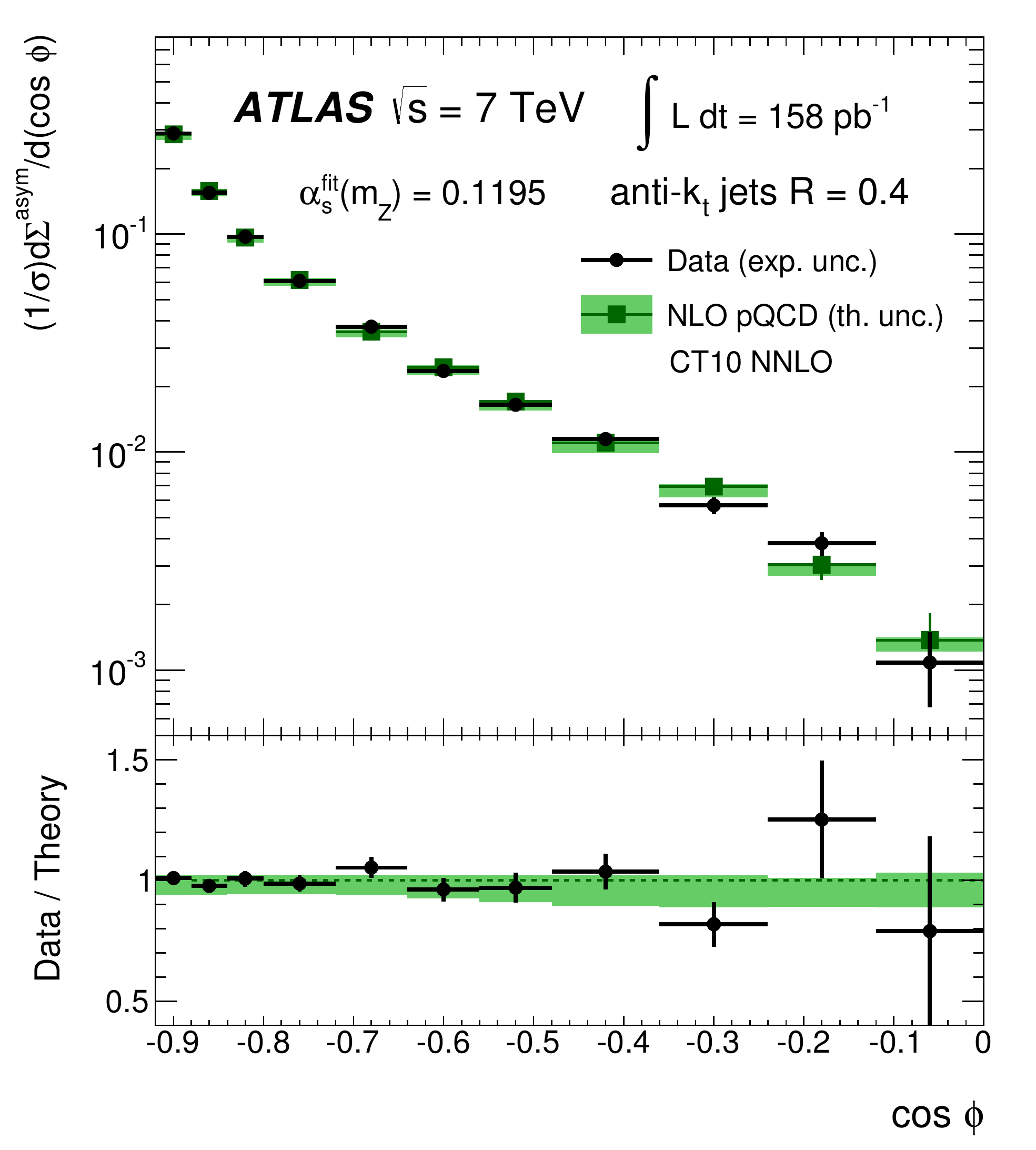}
\caption{The unfolded distributions for transverse energy--energy correlation (left) and its asymmetry (right) compared with the results of a fit to pQCD NLO calculations including non-perturbative corrections. The green shaded band indicates the uncertainty on the theoretical predictions, which includes the sum in quadrature of uncertainties associated with scale, $\alpha_s$, PDF and NPC. The statistical uncertainties on the predictions are indicated by green error bars, appreciable only on the tail of the ATEEC. The solid error bars on the data points (in black) indicate the experimental uncertainties taking into account the correlations between them. The fitted values of the strong coupling constant are $\alpha^{\mathrm{fit}}_{\mathrm{s}}(m_\Zboson) = 0.1173$ (TEEC) and $\alpha^{\mathrm{fit}}_{\mathrm{s}}(m_\Zboson) = 0.1195$ (ATEEC).}
\label{fig:nloComp}
\end{figure}

The theoretical uncertainties on the predictions are treated by varying the 
theoretical distributions by each independent source of uncertainty (scale, 
all independent PDF uncertainties and non-perturbative corrections) and 
repeating the fit using the modified theoretical input.\\

The fit to the TEEC data exhibits shifts in a few nuisance 
parameters, which are always compatible with the $\pm 1\sigma$ band. The 
results for the strong coupling constant obtained using different parameterisations 
of the PDF are summarised in Table \ref{tab:alphaSEEC}, together with the 
experimental uncertainties and the values of $\chi^2/N_{\mathrm{dof}}$.
\begin{table}[h]
\small
\begin{center}
\begin{tabular}{ccc}
\hline
\T\B PDF & $\alpha_{\mathrm{s}}(m_\Zboson)$ value & $\chi^2/N_{\mathrm{dof}}$\\
\hline
\TB\B MSTW 2008 & $0.1175 \pm 0.0010 \mbox{ (exp.) } ^{+0.0059}_{-0.0019} \mbox{ (scale) } \pm 0.0006 \mbox{ (PDF) } \pm 0.0002 \mbox{ (NPC) }$ & 29.0 / 21\\
\TA\B CT10 & $0.1173 \pm 0.0010 \mbox{ (exp.) } ^{+0.0063}_{-0.0020} \mbox{ (scale) } \pm 0.0017 \mbox{ (PDF) } \pm 0.0002 \mbox{ (NPC) }$ & 28.4 / 21\\
\TA\B NNPDF 2.3 & $0.1183 \pm 0.0010 \mbox{ (exp.) } ^{+0.0059}_{-0.0013} \mbox{ (scale) } \pm 0.0009 \mbox{ (PDF) } \pm 0.0002 \mbox{ (NPC) }$ & 29.3 / 21\\
\TA\B HERAPDF 1.5 & $0.1167 \pm 0.0007 \mbox{ (exp.) } ^{+0.0040}_{-0.0008} \mbox{ (scale) }^{+0.0007}_{-0.0024} \mbox{ (PDF) }\pm 0.0001 \mbox{ (NPC) }$ & 28.7 / 21\\
\hline
\end{tabular}
\end{center}
\caption{Results for $\alpha_{\mathrm{s}}$ from fits to the TEEC function using different PDF sets, namely MSTW 2008, CT10, 
NNPDF 2.3 and HERAPDF 1.5, together with experimental as well as theoretical uncertainties
due to scale and PDF choices and non-perturbative corrections.}
\label{tab:alphaSEEC}
\end{table}

The final value for the TEEC fits is chosen to be the one obtained using \textsc{CT10}, 
since its PDF uncertainty is largest and serves as an envelope covering the variations with different 
PDF sets as shown in Table \ref{tab:alphaSEEC}:
\begin{equation}
\alpha_{\mathrm{s}}(m_\Zboson) =  0.1173 \pm 0.0010 \mbox{ (exp.) } ^{+0.0063}_{-0.0020} \mbox{ (scale) } \pm 0.0017 \mbox{ (PDF) } \pm 0.0002 \mbox{ (NPC) }.
\end{equation}

The fit to the ATEEC data does not show any significant shift in the values of the nuisance parameters. In this case, the fit results 
in the values for the strong coupling constant which are summarised in Table \ref{tab:alphaSAEEC}.

\begin{table}[h]
\small
\begin{center}
\begin{tabular}{ccc}
\hline
\T\B PDF & $\alpha_{\mathrm{s}}(m_\Zboson)$ value & $\chi^2/N_{\mathrm{dof}}$\\
\hline
\TB\B MSTW 2008 & $0.1195 \pm 0.0017 \mbox{ (exp.) } ^{+0.0055}_{-0.0015} \mbox{ (scale) } \pm 0.0006 \mbox{ (PDF) }$ & 12.7 / 10\\
\TA\B CT10 & $0.1195 \pm 0.0018 \mbox{ (exp.) } ^{+0.0060}_{-0.0015} \mbox{ (scale) } \pm 0.0016 \mbox{ (PDF) }$ & 12.6 / 10\\
\TA\B NNPDF 2.3 & $0.1206 \pm 0.0018 \mbox{ (exp.) } ^{+0.0057}_{-0.0013} \mbox{ (scale) } \pm 0.0009 \mbox{ (PDF) }$ & 12.2 / 10\\
\TA\B HERAPDF 1.5 & $0.1182 \pm 0.0013 \mbox{ (exp.) } ^{+0.0041}_{-0.0008} \mbox{ (scale) }^{+0.0007}_{-0.0025} \mbox{ (PDF) }$ & 12.1 / 10\\
\hline
\end{tabular}
\end{center}
\caption{Results for $\alpha_{\mathrm{s}}$ from fits to the ATEEC function using different PDF sets, namely MSTW 2008, CT10, 
NNPDF 2.3 and HERAPDF 1.5, together with experimental as well as theoretical uncertainties
due to scale and PDF choices. The uncertainty due to the non-perturbative corrections is negligible.}
\label{tab:alphaSAEEC}
\end{table}
The final value for the ATEEC fit is also chosen to be the one obtained using the \textsc{CT10} parton distribution functions:
\begin{equation}
\alpha_{\mathrm{s}}(m_\Zboson) =  0.1195 \pm 0.0018 \mbox{ (exp.) } ^{+0.0060}_{-0.0015} \mbox{ (scale) } \pm 0.0016 \mbox{ (PDF)}.
\end{equation}

The agreement between the fitted theoretical NLO predictions, including non-perturbative corrections,
and the data is good as shown in Fig.~\ref{fig:nloComp} and indicated by the 
$\chi^2$ values given in Tables \ref{tab:alphaSEEC} 
and \ref{tab:alphaSAEEC}. 
Restricting the angular region in the fits to $(-0.72, 0.72)$, yield
values of the strong coupling constant which vary within experimental uncertainties.
The values of $\alpha_{\mathrm{s}}(m_\Zboson)$ 
found in this analysis are in agreement with the world average 
$\alpha_{\mathrm{s}}(m_\Zboson) = 0.1185 \pm 0.0006$ \cite{pdg}, as well as with other 
determinations of the strong coupling constant from the data collected at the 
LHC \cite{cmsAs,cms3jet,bogdan}.

Calculations beyond NLO accuracy, which are already
available for processes such as top-quark pair \cite{kidonakis} or Higgs boson
production \cite{florian}, are needed for multi-jet production at LHC energies.
They are expected to reduce the scale uncertainties, which are the limiting
factor in this determination of the strong coupling constant.

\section{Summary}
\label{sec:conclusions}
First measurements of the TEEC and ATEEC functions are presented using 158 $\ipb$ of
$pp$ collision data at 7 $\TeV$ recorded by the ATLAS experiment at the LHC.
For this purpose, multi-jet final states are selected requiring jets,
reconstructed with the anti-$k_t$ algorithm and radius parameter $R=0.4$,
with $\pt>50~\GeV$ and $|\eta|<2.5$ and such that the scalar sum of the
transverse momenta of the two leading jets is above $500 \GeV$. 
The TEEC and ATEEC data are fairly well described by \textsc{Pythia 6} and \textsc{Alpgen}, while the \textsc{Herwig++}
MC simulation shows some discrepancies which can be as large as 30\%.\\
The TEEC and the ATEEC at the particle level are compared to perturbative QCD predictions
at NLO accuracy. The renormalisation and factorisation scales are chosen to be $(p_{\mathrm{T}1} + p_{\mathrm{T}2})/2$,
ranging from $250$ to $1300 \GeV$ and with an average value of $305 \GeV$.
Through their construction, both the TEEC and ATEEC functions are
less affected by experimental effects such as the jet energy scale and
resolution or pileup than absolute cross-section measurements.
Similarly, the PDF uncertainties in their theoretical predictions, as given by Eq.~(\ref{eq:TEEC-QCD}), cancel to a large extent.
This renders these observables well suited to determine the strong coupling constant. The data for
$|\cos\phi| < 0.92$ are fitted to the QCD predictions obtained
with \textsc{NLOJet++} to determine the value of the strong coupling
constant. For the TEEC, which provides the experimentally more accurate determination, the result of the fit using the \textsc{CT10} PDF yields

\begin{equation}
\alpha_{\mathrm{s}}(m_\Zboson) =  0.1173 \pm 0.0010 \mbox{ (exp.) } ^{+0.0063}_{-0.0020} \mbox{ (scale) } \pm 0.0017 \mbox{ (PDF) } \pm 0.0002 \mbox{ (NPC)}.
\end{equation}
The present determination of $\alpha_{\mathrm{s}}(m_\Zboson)$ is limited by the uncertainties due to the choice of
renormalisation and factorisation scales.

\paragraph{\bf{Acknowledgements}}{
We thank CERN for the very successful operation of the LHC, as well as the
support staff from our institutions without whom ATLAS could not be
operated efficiently.\\
We acknowledge the support of ANPCyT, Argentina; YerPhI, Armenia; ARC,
Australia; BMWFW and FWF, Austria; ANAS, Azerbaijan; SSTC, Belarus; CNPq and FAPESP,
Brazil; NSERC, NRC and CFI, Canada; CERN; CONICYT, Chile; CAS, MOST and NSFC,
China; COLCIENCIAS, Colombia; MSMT CR, MPO CR and VSC CR, Czech Republic;
DNRF, DNSRC and Lundbeck Foundation, Denmark; EPLANET, ERC and NSRF, European Union;
IN2P3-CNRS, CEA-DSM/IRFU, France; GNSF, Georgia; BMBF, DFG, HGF, MPG and AvH
Foundation, Germany; GSRT and NSRF, Greece; RGC, Hong Kong SAR, China; ISF, MINERVA, GIF, I-CORE and Benoziyo Center, Israel; INFN, Italy; MEXT and JSPS, Japan; CNRST, Morocco; FOM and NWO, Netherlands; BRF and RCN, Norway; MNiSW and NCN, Poland; GRICES and FCT, Portugal; MNE/IFA, Romania; MES of Russia and NRC KI, Russian Federation; JINR; MSTD,
Serbia; MSSR, Slovakia; ARRS and MIZ\v{S}, Slovenia; DST/NRF, South Africa;
MINECO, Spain; SRC and Wallenberg Foundation, Sweden; SER, SNSF and Cantons of
Bern and Geneva, Switzerland; NSC, Taiwan; TAEK, Turkey; STFC, the Royal
Society and Leverhulme Trust, United Kingdom; DOE and NSF, United States of
America.\\
The crucial computing support from all WLCG partners is acknowledged
gratefully, in particular from CERN and the ATLAS Tier-1 facilities at
TRIUMF (Canada), NDGF (Denmark, Norway, Sweden), CC-IN2P3 (France),
KIT/GridKA (Germany), INFN-CNAF (Italy), NL-T1 (Netherlands), PIC (Spain),
ASGC (Taiwan), RAL (UK) and BNL (USA) and in the Tier-2 facilities
worldwide.
}

\clearpage

\bibliographystyle{atlasBibStyleWithTitle}
\bibliography{eecPaper}

\newpage \begin{flushleft}
{\Large The ATLAS Collaboration}

\bigskip

G.~Aad$^{\rm 85}$,
B.~Abbott$^{\rm 113}$,
J.~Abdallah$^{\rm 151}$,
O.~Abdinov$^{\rm 11}$,
R.~Aben$^{\rm 107}$,
M.~Abolins$^{\rm 90}$,
O.S.~AbouZeid$^{\rm 158}$,
H.~Abramowicz$^{\rm 153}$,
H.~Abreu$^{\rm 152}$,
R.~Abreu$^{\rm 116}$
% ...
\bigskip
\\
% ...
$^{11}$ Institute of Physics, Azerbaijan Academy of Sciences, Baku, Azerbaijan\\
% ...
$^{85}$ CPPM, Aix-Marseille Universit{\'e} and CNRS/IN2P3, Marseille, France\\
% ...
$^{90}$ Department of Physics and Astronomy, Michigan State University, East Lansing MI, United States of America\\
% ...
$^{107}$ Nikhef National Institute for Subatomic Physics and University of Amsterdam, Amsterdam, Netherlands\\
% ...
$^{113}$ Homer L. Dodge Department of Physics and Astronomy, University of Oklahoma, Norman OK, United States of America\\
% ...
$^{116}$ Center for High Energy Physics, University of Oregon, Eugene OR, United States of America\\
% ...
$^{151}$ Institute of Physics, Academia Sinica, Taipei, Taiwan\\
$^{152}$ Department of Physics, Technion: Israel Institute of Technology, Haifa, Israel\\
$^{153}$ Raymond and Beverly Sackler School of Physics and Astronomy, Tel Aviv University, Tel Aviv, Israel\\
% ...
$^{158}$ Department of Physics, University of Toronto, Toronto ON, Canada\\
% ...
\end{flushleft}

\end{document}